\input epsf
\documentclass[nohyper,notoc]{article} 
\usepackage{epsfig}

\def\beq{\begin{equation}}
\def\eeq{\end{equation}}
\def\bea{\begin{eqnarray}}
\def\eea{\end{eqnarray}}
\def\bq{\begin{quote}}
\def\eq{\end{quote}}
\def \lsim{\mathrel{\vcenter
     {\hbox{$<$}\nointerlineskip\hbox{$\sim$}}}}
\def \gsim{\mathrel{\vcenter
     {\hbox{$>$}\nointerlineskip\hbox{$\sim$}}}}
\def\gappeq{\mathrel{\rlap {\raise.5ex\hbox{$>$}}
{\lower.5ex\hbox{$\sim$}}}}
\def\lappeq{\mathrel{\rlap{\raise.5ex\hbox{$<$}}
{\lower.5ex\hbox{$\sim$}}}}

\def\meg{\mu \to e \gamma}

\def\m3e{\mu \to e \bar{e} e}
\def\a{\alpha}
\def\b{\beta}

\def\m{\mu}

\begin{document}
\renewcommand{\thefootnote}{\fnsymbol{footnote}}
\begin{center}
{\Large {\bf 
  Similar Dark Matter and Baryon abundances
with  TeV-scale  Leptogenesis  
}}
\vskip 25pt
{\bf   Sacha Davidson $^{1,}$\footnote{E-mail address:
s.davidson@ipnl.in2p3.fr} and  Martin Elmer $^{1,}$}\footnote{E-mail address:
m.elmer@ipnl.in2p3.fr} 
 
\vskip 10pt  
$^1${\it IPNL, Universit\'e de Lyon, Universit\'e Lyon 1, 
CNRS/IN2P3, 4 rue E. Fermi 69622 Villeurbanne cedex, France} \\
\vskip 20pt
{\bf Abstract}
\end{center}
We estimate the Baryon Asymmetry of the Universe (BAU) produced
in an inverse seesaw model containing extra light singlets,
and with lepton number conservation prior to the electroweak phase
transition.  A CP asymmetry $\epsilon \sim
{\cal O}(1)$ is required to obtain a large enough BAU.
We discuss the relation between 
the baryon and WIMP relic densities in
baryogenesis scenarios  using the
out-of-equilibrium
decay of a baryon-parent of mass $M$
  : when baryon number violation
freezes out,  the remaining density
of baryon-parents   is $\sim M/m_W \times$ the WIMP relic
density. So the baryon/WIMP ratio is  $\sim \epsilon M/m_W$.
A natural explanation of the similar WIMP and baryon densities could
be that CP violation is of order the ratio $m_W/M$.

\begin{quotation}
  {\noindent\small 

\vskip 10pt
\noindent
}

\end{quotation}

\vskip 20pt  

\setcounter{footnote}{0}
\renewcommand{\thefootnote}{\arabic{footnote}}

\section{Introduction}
\label{intro}

In recent years, there
has been some interest in relating the cosmological  
number density of baryons\cite{Sakharov,BAU,PRep},  with the number density of
dark matter particles\cite{DM}. A popular approach, refered to
as Asymmetric Dark Matter (ADM)\cite{ADMtot,ADM,ADMl},   is to
implement a single mechanism that generates an
excess of 
particles over anti-particles, which materialises both in 
the  baryons and in the  dark
matter. However, 
it is not necessary to relate
the  amounts of Dark Matter and baryons, but
rather, to explain that  both have their observed values. This
(subtle) distinction is relevant, because
the relic  density of  a weakly interacting
particle (WIMP)  is  naturally of
order the observed dark matter density. So  it is not
clear  what is gained  by dropping the ``WIMP miracle'', to
link the   dark matter
number density   to the baryon asymmetry,  which depends
on an arbitrary  CP violation parameter. 
Scenarios which implement the link in the
inverse direction \cite{CRS,JM}, explaining the  baryon
number density  from  the  dark matter  density,
could be more interesting. 
But they  require a CP asymmetry  
that is naturally ${\cal O} (1)$.
The aim of this paper is to  
follow the  ``subtle'' distinction mentioned above:
we assume that the Dark Matter are WIMPs (so their
observed relic abundance is already  natural and requires
no explanation), and attempt to build a baryogenesis model,
which naturally generates, in a similar cosmology,
the observed baryon asymmetry.

We consider an  ``inverse seesaw''-like  model \cite{inverse},
extending the Standard Model(SM) with  TeV-scale
 electroweak singlet Dirac neutrinos $\psi$, 
 additional light singlet fermions $s$
and a  $L=2$ singlet scalar $\phi$. 
The $\psi $ participate in  Yukawa interactions with SM
 neutrinos, and the  $\phi$  have
($L$ conserving) interactactions with $\psi$ and $s$.
For generic choices of couplings, 
lepton number
violation  arises   spontaneously
when $\phi$ gets a vev,  after
the electroweak phase transition,
 giving Majorana masses
to the doublet SM  neutrinos. However,
there is a particular limit where
lepton number is conserved and the SM neutrinos
are Dirac, whose baryogenesis prospects
are explored in section \ref{deME}.
Section  \ref{notation} reviews our model and
notation. 
 Since we are building a model,
we would like it to exhibit as many interesting features
as possible. Our model  can fit the observed neutrino
mass differences. However,  the  additional light  singlets
cannot fit the reactor neutrino  anomaly\cite{neu} and Big
Bang Nucleosynthesis;
this is briefly discussed in section  \ref{notation}.

The Baryon Asymmetry of the Universe (BAU)
is generated via leptogenesis \cite{FY}
at the TeV scale.
As the temperature of the Universe drops
below the mass of the singlet $\psi$s, their number density
is depleted by annihilations and decays. 
We require  this to occur prior to the
electroweak phase transition.
 They can  decay 
to  SM neutrinos $\nu$   and a Higgs, or to light singlets $s$  and $\phi$,
and CP asymmetries can arise in these decays.
Since lepton number is conserved until $\phi$ gets
a vev,  these
lepton asymmetries would be  of equal magnitude and opposite sign in
 the doublets
and singlets.
The asymmetries can survive in the plasma, once
``washout interactions'', exchanging
lepton number between the doublets
and singlets, are out of equilibrium.
Section \ref{sec:equil} estimates  that
the co-moving number density  of $\psi$s remaining,
when this occurs, is ``naturally'' of order that of WIMPs.
So to  ``naturally'' 
obtain similar  relic
densities of baryons and WIMPs would require
to ``naturally'' obtain ${\cal O}(1)$ CP asymmetries
in the $\psi$ decays.
As reviewed   in section \ref{sec:CP},
 large  CP asymetries  can be arranged  
by taking the $\psi$ masses 
to be quasi-degenerate\cite{Pilaftsis}. 
The lepton
asymmetry in Standard Model neutrinos
then  can be partially transformed  to  baryons
by non-perturbative SM  B+L violation.  Dark matter is
assumed to be some other WIMP, with a ``usual'' relic
abundance. 
So this scenario gives similar abundances of
WIMPs and baryon-{\it parents}, but
similar    dark matter and
baryon number densities only  arise 
if there is an ${\cal O}(1) $  
CP asymmetry  in the parent  decays.
 Unfortunately,
such a large CP asymmetry is not easy to obtain
in our model. 
We comment on this scenarios ability to  relate  
the baryon and 
dark matter densities in section \ref{disc}.


\section{Review,  Notation and Masses}
\label{notation}

\subsection{Observations}

The mass density of baryons $B$ in the Universe
today, as extracted from a canonical  $\Lambda CDM$
model \cite{WMAP}, is 
\beq
 \frac{m_p( n_B - n_{\overline B})}{\rho_c}= 
 \Omega_{B} h^2 = 0.02255 \pm 0.0054 ~~~~~~\Rightarrow
~~~~~~~~Y_B \equiv  \frac{( n_B - n_{\overline B})}{s} \simeq 
8.7 \times 10^{-11} 
\label{YB}
\eeq
where $s = 2 \pi^2 g_* T^3/45$ is the
entropy density,
 $\rho_c = 3H_0^2/(8 \pi G)$ is the critical density
to obtain a spatially flat   Universe, and 
 $H_0 = 74.2 \pm 3.6$ km/sec/Mpc  $= 100 h $  km/sec/Mpc.
The cold dark matter relic abundance, for the same 
 canonical $\Lambda CDM$ cosmology with three light neutrinos,
is 
\beq
\Omega_{DM} h^2 = 0.1126\pm 0.0036
\eeq
so that the ratio of dark matter particles to baryons
is 
\beq
\label{rapport}
\frac{Y_{DM}}{Y_{B}} \sim 5 \frac{m_p}{m_{DM}} ~~,
\eeq
where $m_p$ is the proton mass.
Since $m_p$ mostly  arises from QCD, 
and the dark matter should not have strong
interactions, we assume that $m_p$ and $m_{DM}$
are unrelated, and focus on obtaining similar
baryon and dark matter number densities today:
$  Y_{B}/30  \lsim  Y_{DM} \lsim 30 Y_{B}$.
This  definition of ``similar'' allows a dark matter 
candidate in the mass range accessible to most
direct detection experiments.

 Notice that the ratio (\ref{rapport}) could be
different  in more
compliciated cosmological models, such as those with
extra sterile neutrinos. The CMB data
alone allows the dark matter density to
increase with the number of relativistic species
present at recombination (counted as
the number $N_{eff}$ of neutrino flavours):
$\delta \Omega_{DM} \sim .4 ( \delta N_{eff} \Omega_{DM}/N_{eff})$
\cite{WMAP}. Including other observations,
such as $H_0$, favours  $ N_{eff}  =  4.34^{+0.86}_{-0.88} $(68$\%$ CL) 
 \cite{WMAP,GGR}.
An increase in $\Omega_{DM}$ would usefully
increase the ratio (\ref{rapport}),   various
laboratory neutrino anomalies\cite{neu}
 could benefit from light neutrinos
other than the three of the Standard Model,
and this paper constructs a neutrino mass
model with  additional light singlets. 
However, cosmological data also constrains
the light neutrino mass scale.  An additional
eV sterile neutrino that could address the
reactor neutrino anomaly should respect
cosmological  bounds on hot dark matter. 
It was shown by \cite{GGR} that such a
eV-sterile is cosmologically
consistent if there are additional
lighter neutrino species to delay
matter radiation equality, and
an  ${\cal O}(0.01)$ asymmetry in
the $\nu_e$ density to allow
 Big Bang Nucleosynthesis (BBN) \cite{BBN,MS2} 
to produce 
the observed  relic abundances
of light elements. Our model allows
several sterile species, but the
baryon asymmetry is related to the asymmetry
in SM neutrinos, so  we do not obtain
the large $\nu_e$ chemical potential 
required by \cite{GGR,MS2}. Furthermore,
if our extra  steriles decouple
prior to the Electroweak Phase Transition
(EPT), their contribution to the
radiation density  at BBN is suppressed by
a factor $g_{*}(T_{BBN})/g_{*}(T_{EPT})$,
where $g_{*}(T)$ is the number of
Standard Model  degrees
of freedom   in the plasma at $T$. 
 
\subsection{Models}
\label{sec:models}

An  explanation for the  ratio
(\ref{rapport}), which has benefited
from recent interest\cite{ADM,ADMl} 
  could be that  a single asymmetry
controls both the baryon and dark matter relic
number densities.  
If this ratio, given in eqn (\ref{rapport}), should be exactly one,
then a comparatively light
cold dark matter candidate is required. 
A model where the DM gets a mass after
the EPT is \cite{OS}.
The observed difference
in  mass densitites could also  be due to
 inequivalent 
cosmological histories after the formation of
the asymmetry.   Some early discussions of
asymmetries in the dark matter density are
\cite{ADMtot,FZ};  for   a review of recent  Asymmetric
Dark Matter (ADM) scenarios, see {\it e.g.} \cite{ADMrev}.
Asymmetric leptogenesis models
\cite{ADMl}, and   links to Dark Energy
and inflation \cite{Guo,AGS}
 also have been explored.
ADM  models explain why the dark matter and
baryon number densities are similar, but not
why they have have the observed value.

An alternative approach \cite{CRS,JM}, 
attaches importance to the  `` WIMP miracle''
(see {\it eg} \cite{DM}):
the relic abundance   of 
 a massive particle,  which annihilates with
its anti-particle via a weak cross-section, 
gives $\Omega_{DM} h^2 $  of order the  observed value. 
To explain the ratio
(\ref{rapport}), one  then only needs a natural
scenario giving a relic  number density   of baryons
minus anti-baryons,  which is 
 similar to the relic  number   density of
WIMPS plus anti-WIMPS. It is
well-known
\cite{Sakharov}  that to 
produce an asymmetry,  CP violation
and non-equilibrium  are required,
whereas
only a departure from thermal equilibrium is required for 
 the particle-antiparticle-symmetric relic density of WIMPs. 
So  if the non-equilibrium is similar
in the baryon and dark matter production,
and  the CP asymmetry for baryogenesis is one, then
similar number densities could be expected.
For instance,
in \cite{CRS}, the baryon asymmetry is
produced in the out-of equilibrium
annihilations of the dark matter particle,
which works for selected mass
ranges of the particles involved.
In this paper,  the lepton asymmetry is
produced in decays. 
The parent particle of the asymmetry
first annihilates, then decays out-of-equilibrium,
at the TeV-scale.
 This freezes-in the baryon-parent
and dark matter densities at a similar
temperature,  so the  departures from
thermal equilibrium in both processes
may be similar.  
Indeed, the co-moving 
density of a weak-scale 
unstable particle, when inverse decays freeze out,
is similar to the relic density of a weak-scale
particle that annihilates.
So  an  ${\cal O} (1)$ CP asymmetry
is required   to get  similar baryon and
dark matter   densities today.

We construct  a  TeV-scale model of out-of-equilibrium-decay  leptogenesis
\cite{FY}.
Various mechanisms
 have been studied for leptogenesis
\footnote{
Low temperature out-of-equilibrium-decay baryogenesis
scenarios have also been considered; the 
analysis of \cite{CHH} was useful to us.}
 in
TeV-scale seesaw models \cite{TeVLH},
such as degenerate decaying singlets \cite{Pilaftsis},
or  adding extra particles \cite{TeVL}.  Heavy
singlet fermions in the ``inverse seesaw'' pattern
have been studied by various people \cite{CJN,TeVLinvH,FXM,TeVLinv}.
Our model differs  from the usual inverse
seesaw  in that we have additional light singlets\cite{Zhang},
so we can generate a non-zero  asymmetry
in SM leptons, despite that the model conserves
lepton number. Our leptogenesis scenario therefore  differs
from that of  \cite{CJN}, who also study a lepton number
conserving scenario, but without extra light singlets. 
They obtain a sufficient baryon asymmetry by
turning off the sphalerons before inverse
decays go out of equilibrium, when various asymmetries
are present and changing. Here, we explore  whether
the baryon asymmetry can  be obtained  in a more
``dynamics-independent'' way, due to the presence of
extra singlets.  
The Lagrangians of \cite{TeVLinvH} contain
``hard'' lepton number breaking, so the relevant
interactions for leptogenesis are somewhat different,
because a net lepton asymmetry in generated. 
An advantage of the hard breaking is that 
it can naturally provide the small mass splitting
between the heavy decaying singlets\cite{TeVLinvH}.
The model of \cite{FXM} generates the asymmetry in
the TeV-scale singlets.
We focus on a lepton number conserving model,
because  it is for such models that a ``hidden
sector'' (as provided by the  extra light singlets)
is useful for storing an asymmetry\cite{DW}.
Futhermore,  the natural   scale of
lepton number violation in the inverse seesaw
is below the electroweak scale, so if lepton number
violation is spontaneous,   one could
expect it to arise after the electroweak
phase transition.

The inverse seesaw \cite{inverse} has received attention
because it generates small neutrino masses 
due to  new  particles with TeV-scale masses and
couplings of a ``natural'' size.  Such  new particles
may therefore be kinematically accessible
to the LHC \cite{NLHC}, or induce detectable \cite{nUmeg}
non-unitarity and/or Lepton Flavour Violation.
We wish to explore,
in models  where the small lepton number violation
is spontaneous,  the prospects of
additional light  singlets and leptogenesis.
 A careful study \cite{CJN} (without additional
light singlets $s$), showed
that the baryon asymmetry can be generated by
turning off the sphalerons during the  decays.
We had hoped that baryogenesis would be easier
with additional light singlets, but the parameter
space where we  estimate that it could work is
not large enough to be convincing. 

Dark Matter has been studied in radiative inverse
seesaw models \cite{RadInvSee1,RadInvSee2}, where it can
arise more
naturally than  in the inverse seesaw \cite{DMinverse}. 
We recall that
our model does not aim to provide a WIMP;
we suppose that it arises in some other sector
of the theory. 

\subsection{Notation and Masses}
\label{sec:mnu}

We consider a   Lagrangian with several additional singlets,
which can spontaneously violate   lepton number. 
The SM  doublet leptons  $\ell$ have Yukawa couplings
to (two or three) `` right-handed neutrinos''
$N_R$,  there is some larger
number of   singlet left-handed leptons  $\{ S,s \}$,
and  an $L = -2$ singlet
scalar $\phi$.
The left-handed singlets who participate 
in a mass matrix $M$ with the $N_R$ 
are refered to as $\{ S \}$,   and
the remaining $s$s can acquire masses
once $\phi$ develops a vev.
Allowing all  renormalisable interactions, 
the Lagrangian, 
written in two-component notation with
all the fermions left-handed, is:
\beq \label{L} 
{\cal L}
 = 
{\cal L}_{SM} +
 \{
{  \lambda^*} [\ell \tilde{H}]  N^c + 
{ M^*}\, N^c  S + { y^*} \,\phi S s + \frac{{ Y^*}}{2} \phi S S + 
\frac{{ X^*}}{2} \phi s s 
+\frac{{ Z^*}}{2} \phi^\dagger   N^c N^c
+ h.c. \}  + V(\phi)~~,
\eeq
$ \tilde{H}$ is the Higgs,  
$ [\ell \tilde{H}] = \nu \tilde{H}_0 - e\tilde{H}_+ $,
and
$N^c, S$ and $s$
have  lepton numbers $L
=\{-1,1,1\}$. 
 The potential $V(\phi)$
causes $\phi$ to develop a vev 
after the electroweak phase transition. 
We assume that such a potential can be constructed,
and consider it no further. We
also ignore the resulting majoron
\cite{majoron}. 
Generation indices are implicit;
for simplicity, we suppose two generations of $N, S$ and $s$, so
the  matrices $ { \lambda, y, Y, X, Z}$ and ${ M}$
are two by two. We denote the eigenvalues of
$\lambda$ as $\lambda_1, \lambda_2$ (and similarly for
other  matrices). We will see that 
 $\lambda_1 \sim \lambda_2$, $y_1 \sim y_2$ and
 $M_1 \sim M_2$ to obtain a large enough  CP
asymmetry for leptogenesis.

 Prior to the
Electroweak Phase Transition,
 there are two 
gauge singlet Dirac fermions, $\psi_1, \psi_2$:
\beq
\psi_I = \left(
\begin{array}{c}
S_I \\
N_I
\end{array} \right) ~~,
\label{psi}
\eeq
 whose masses $M_I \sim $ TeV  are
the eigenvalues of $M$.
 The $M_I$
will be taken degenerate to obtain a large
enough baryon asymmetry, so when the mass
difference is irrelevant, we write  $M_\psi$.


In the presence of vacuum expectation values
$v = \langle H \rangle = 174$ GeV and $u = \langle \phi \rangle \lsim v$, 
the neutral
Majorana mass matrix can be written, in two component notation, as
\beq
\label{mass}
{\cal L}_{mass} = -\frac{1}{2} \left( \nu_L \, N^c \,  S \,  s \right)
\left[
\begin{array}{cccc}
 0& m_D &0&0  \\
 m^T_D &\mu_Z&M &0 \\
0& M^T &\mu_Y& \mu_y\\
0& 0  &\mu_y^T & \mu_X 
\end{array}
\right]
\left(
\begin{array}{c}
   \nu_L \\ N^c \\ S \\ s 
\end{array}
\right) + h.c. 
\eeq
where $
m_D  =  \lambda v $,
$\mu_X =  X u$,
$\mu_y =  y u$,
$\mu_Z =  Z u$,
and $\mu_Y =  Y u$ are two by two matrices.

We are interested in the limit where the eigenvalues of $M$
are much larger than the other entries in the mass matrix. In the case
of the usual
inverse seesaw, which  contains no light singlets $s$,   
the determinant  implies that the light
active neutrino mass matrix is  $\sim m_D M^{-1} \mu_Y  M^{T -1} m_D^T$.
In the case where  extra singlets are present, the determinant  of
the mass matrix in (\ref{mass}), in one generation,
 is  $ m_D^2(\mu_Y \mu_X - \mu_y ^2)$. 
We neglect from now on the coupling $Z$, because its
contributions to the mass matrix are unimportant,
and its effects in  the  baryogenesis scenario
are similar to those of $Y$, which will be
unimportant.
We will see in section \ref{sec:equil} that  the couplings
 in $y $ and/or
$\lambda$ must be small\footnote{
We therefore do not expect bounds on
our model from  lepton flavour violating processes
such as $\meg$, or  non-unitarity \cite{nUmeg}.} ($\lsim 10^{-5}$),
to prevent the washout of the lepton asymmetry. We focus on
 two simple  limiting cases:  $y_i \ll Y_i, X_i \lsim 1$ , which gives
Majorana
masses for the active neutrinos, and the case 
$ Y_i =  X_i = 0$, which
gives Dirac masses between the $\nu_L$ and $s$.

In the Majorana case with  $y_i \ll Y_i, X_i$,  the active neutrinos
have a usual-inverse-seesaw-like  mass matrix
$m_\nu \simeq  m_D M^{-1} \mu_Y  M^{T -1} m_D^T$,
whose eigenvalues satisfy
\beq
\frac{m_{atm}m_{sol}}{{\rm eV}^2}
= 0.3
\left( \frac{\lambda_1 \lambda_2}{10^{-9}}\right)^2
\left( \frac{(3 ~{\rm TeV})^2}{M_1 M_2}\right)^2
  \frac{Y_1 Y_2 \langle \phi \rangle^2}{v^2}
\label{mnunum}
\eeq
   $\phi$ should  have a 
mass and vev well below  the Higgs vev $v =174 $ GeV, to ensure that lepton
number is conserved prior to the electroweak
phase transition. However,
we will be forced to take  the vev of order $v$
(to allow  $\lambda_i$ small enough
to give a sufficient abundance of $\psi$s
after freezeout of washout interactions).
We will also require $\lambda_1 \simeq \lambda_2$,
to obtain a large enough CP asymmetry, so
we will suppose a mild hierarchy in $Y$ to
generate the atmospheric - solar splitting.

The light 
steriles $s$  have a mass matrix
$\sim \mu_X = X \langle \phi \rangle$. In
the one generation case, $s$ has a  mixing angle with
the active neutrino of order 
$ \mu_y m_D/(M \mu_X)$, which allows the
decay $s \to 3 \nu$ at tree level. We focus
on  parameters\footnote{With $X \sim 10^{-10}$,  we can
obtain a singlet $s$ with eV  mass and $\sim 0.1 -0.01$
mixing angle, that could fit the reactor
neutrino anomaly. However, we do not explore this
parameter space, because out model cannot give the
large lepton asymmetry required to make it cosmologically
acceptable.}   $X \lsim Y$,  such that 
the singlets are innocuous: their masses are  $\lsim$ GeV,
they annihilate efficiently (to majorons), and
otherwise can decay. Since the $s$s  decouple
prior to the Electroweak Phase Transition, 
their temperature at $T \sim \mu_X$  is 
suppressed with respect
to the  photons, so they should not
 over-contribute to the radiation density.

The active neutrinos share Dirac masses $\sim 
m_D\mu_y/M$ with the singlets
$s$, if $X = Y = Z = 0$ in the Lagrangian (\ref{L}). 
We  consider this limit only  in section 
\ref{deME}, because it allows  more parameter space for our leptogenesis
scenario\footnote{Mixed scenarios can be envisaged, for instance where
the atmospheric mass is Dirac,  and
the solar mass is Majorana. We do not study this
tuned example because we estimate that
out-of-equilibrium-decay-leptogenesis does
not occur.}. By assigning $L = 0$ to
$\phi$, and -1 to $s$,     
lepton number clearly is conserved also
after both $\phi$ and $H$ get vevs. 
However,  since  $\phi$  was introduced  to 
spontaneously break lepton number, 
it is   peculiar   to not give it
lepton number. We nonetheless
assume that $\phi$ does not mix with $H$,
and do not discuss
potential constraints from Higgs physics.


\section{Thermal History}

One of the roles of $\psi$ is to generate the Baryon Asymmetry
of the Universe (BAU). In its decays, it can
generate an active versus sterile lepton asymmetry  and
the excess of active leptons will be partially transformed
to baryons by sphalerons \cite{sphalerons}. 
This is discussed in section \ref{sec:CP}. The asymmetries
produced in $\psi$ decay  can survive in the plasma once
``washout interactions'' (such as inverse decays
$\ell H \to \psi $ and $\overline{\phi} 
\overline{s} \to \psi $, and scattering 
$\ell H \to \overline{\phi} 
\overline{s}$) go out of equilibrium. This ``freeze-out'' 
temperature $T_{BAU}$ 
is estimated in  section \ref{sec:equil}. The baryon
asymmetry present today will be generated  in the decays
of the $\psi$s remaining at  $T_{BAU}$.

Decays and inverse decays are not 
the only interactions which change the $\psi$ number density;
$ \psi+  \overline{\psi} $ can
also annihilate, via the coupling $Y$ 
to   $ \phi+  \overline{\phi}$. The annihilation
rate is faster than the decay rate
at temperatures just below the $\psi$ mass,
so we start by  discussing  annihilations
 below  in section  \ref{sec:DM}.
Annihilations and inverse decays
 will freeze out at similar  temperatures.

\subsection{$\psi \overline{\psi} \to 
\phi \overline{\phi}$ annihilations}
\label{sec:DM}

At temperatures $T \gsim M_I$, an equilibrium abundance of
$\psi_I$ and $\overline{\psi}_I$ will be present, because
they can be produced via  Yukawa  interactions involving  $\lambda$
(for instance in $ t_R \overline{t_L} \to \tilde{H}^* \to \ell \bar{\psi}$,
at a rate $\Gamma \sim h_t^2 \lambda^2 T/(4 \pi)$, which is fast compared to 
the Universe expansion $H$).

As the temperature $T$ drops below their mass, the
singlets can 
 annihilate via their $Y$ couplings,
or
decay to a light
lepton and a scalar via their $\lambda$
or $y$ couplings:
\bea
\psi_I +\overline{\psi}_J \to \phi + \overline{\phi} \\ 
\psi_I \to \ell + \tilde{H}~ ,
~  \overline{s} +  \overline{\phi}
\eea 
We focus first on  annihilations,
because we envisage $1 \sim Y_i \gg y_j , \lambda_k \lsim 10^{-4}$, 
which implies that
initially, at $T \lsim  M_I$, the $\psi_I$s are more
likely to annihilate than decay.

The non-relativistic annihilation cross-section in the
centre of mass  is
\beq 
\sigma v(\psi_I +\overline{\psi}_J \to \phi + \overline{\phi})
 = \frac{ |[Y Y ^\dagger]_{IJ}|^2}{128 \pi {M}_\psi^2}
\eeq
where all the singlet masses are approximated as ${M}_\psi$. 
We neglect the Sommerfeld enhancement
due to $\phi$ exchange\footnote{We thank Marco Cirelli
for discussions of this issue.} because the $\psi$s are almost 
relativistic; the enhancement can be estimated 
\cite{TDM} to be a few percent.
The number density of $\psi_I$s remaining when 
 annihilations freeze out, and the  
freeze-out temperature  $T_{ann} \equiv M_\psi/z \simeq M_\psi/20$ can be
determined from 
\beq
n_{\psi} (T_{ann})  { \sigma v }
\simeq H(T_{ann}) 
\simeq \frac{1.7
\sqrt{g_{*}(T_{ann})} T_{ann}^2}{m_{pl}}
\label{gelann}
\eeq
where  $g_{*}$ is the 
 number of degrees of freedom in the plasma,
and $g_{*}($TeV) $\simeq  100$ in our model.


\subsection{Decays and the CP asymmetry}
\label{sec:CP}

After the $\psi$ annihilations freeze out, the number
density of $\psi$s and $\overline{\psi}$ will continue
to drop, because they decay. The model conserves lepton
number, so the total lepton asymmetry produced in
these decays is zero. However, since the $\psi$s
decay both to SM particles $H, \ell$, or to
singlets $\overline{s}, \overline{\phi}$,  equal but opposite lepton asymmetries
in the singlet and doublet sectors could be
produced. To obtain a baryon to dark matter
ratio that is ${\cal O}(1)$, these CP
asymmetries will need to be large,
so we explore this limit   below.

\begin{figure}[ht]

\begin{center}
\includegraphics[width=14.5cm]{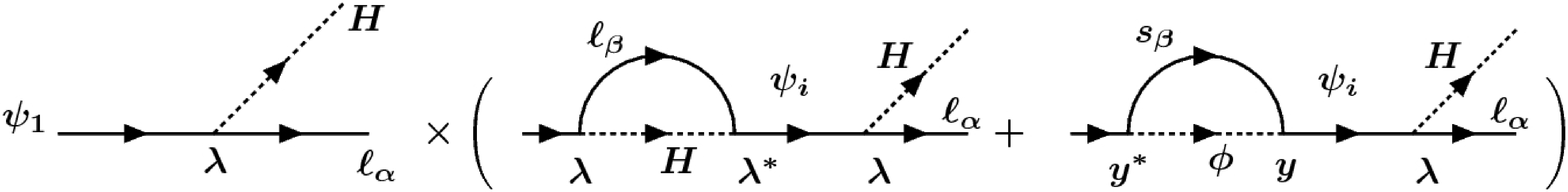}
\end{center}

\caption{ 
diagrams contributing to  the CP asymmetry  with final state $\ell_\a + H$.
\label{fig:lH}} 
\end{figure}

\begin{figure}[ht]
\begin{center}
\includegraphics[width=14.5cm]{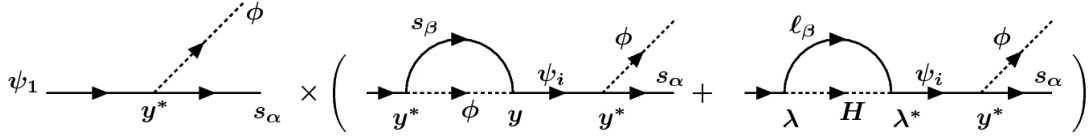}
\end{center}

\caption{ 
 diagrams contributing to  the CP asymmetry
with final state $s_\a + \phi$. 
\label{fig:sphi}} 
\end{figure}

The $\psi_I$s, which carry $L =1$,  can decay at tree level
 to an $\ell_\a$ and an  $H$, or  to  a $\overline{\phi}$
 $(L =2$)
 and  a $\overline{s}_\b$, at partial rates
\beq
\label{Gamtree}
\Gamma(\psi_I \to \ell_\a + H) = 
\frac{|  \lambda|^2_{\a I} }{16\pi}
M_I
~~~,~~~
\Gamma(\psi_I \to \overline{s}_\b + \overline{\phi}) = 
\frac{|y|^2_{\b I} }{32\pi}
M_I
\eeq
See the first diagrams
of figures \ref{fig:lH}  and \ref{fig:sphi}.
These decays are fast, compared to the expansion rate $H$,
for  $T \lsim   \{ |\lambda|, |y|\} \sqrt{M_I m_{pl}}/20$,
so at all temperatures relevant to us.

 CP violating asymmetries for a given final state: 
\beq
\epsilon^{(I, \ell)}_\a = \frac{\Gamma (\psi_I \to \ell_\a H) -
\Gamma (\overline{\psi}_I \to \overline{\ell}_\a \overline{H})}
{\Gamma (\psi_I \to all) +
\Gamma (\overline{\psi}_I \to all)
}
~~~,~~~
\epsilon^{(I, s)}_\a = \frac{\Gamma(\psi_I \to
  \overline{s}_\a \overline{\phi}) -
\Gamma (\overline{\psi}_I \to  s_\a \phi)}
{\Gamma (\psi_I \to all) +
\Gamma (\overline{\psi}_I \to all)
}
\eeq
can be obtained  from the interference of tree 
with loop diagrams, if the couplings are complex and some
particles in the loop can be on-shell. 
In
 figures \ref{fig:lH}  and \ref{fig:sphi}, are
drawn the subset of one-loop diagrams which
we include in our calculation.
We are interested in $\epsilon
\to 1$, which  can be obtained in the decays of
quasi-degenerate singlets  from the illustrated
  wave-function
renormalisation diagrams\cite{Pilaftsis}, so 
 we neglect
the additional  vertex corrections diagrams.

We neglect the effects of $X$ and $Y$ couplings in
the decay rate and CP asymmetries, despite the
hierarchy in magnitudes: $X,Y \sim 1$,
$y, \lambda \sim 10^{-4}$. This is because decays
involving $X$ and $Y$ are suppressed by
 three-body  final state phase space, and we
were unable to find significant effects
of $X$ and $Y$ in the CP asymmetries.

In the decay
of $\psi_I$,
the total asymmetry in the doublets, summed on
lepton flavours $\a$,  is
\beq
\epsilon^{(I, \ell)} = \sum_\a \epsilon^{(I, \ell)}_\a
= 
 \sum_{\a,\b,J} \frac{Im\{\lambda_{\a I} \lambda^*_{\a J } y_{\b I} y^*_{\b J} \}}
{8 \pi [2 \lambda^\dagger \lambda + y^\dagger y]_{II}} \frac{\sqrt{x_J}}{1-x_J}
\label{epsl}
\eeq
where $x_J = M^2_J/M^2_I$. 
This contribution  arises from the interference of the tree
decay with the last loop of figure \ref{fig:lH}. 
Similarly, 
the total asymmetry in the singlets is
\beq
\epsilon^{(I, s)} = \sum_\a \epsilon^{(I, s)}_\a
=
\sum_{\a,\b,J} \frac{Im\{y^*_{\a I} y_{\a J } \lambda_{\b J} \lambda^*_{\b I} \}}
{8 \pi [2 \lambda^\dagger \lambda + y^\dagger y]_{II}} \frac{\sqrt{x_J}}{1-x_J} 
\label{epss}
\eeq

In both the doublet and singlet sectors, there can be  asymmetries
in the individual flavours $\a$, which arise from
the middle diagram of figures \ref{fig:lH} and
\ref{fig:sphi}. We neglect these contributions,
and focus on the flavour-summed singlet and
doublet asymmetries of eqns (\ref{epsl})
and  (\ref{epss}),
because its the
 total  (=flavour-summed) doublet lepton asymmetry
which the sphalerons transform 
to baryons\footnote{The flavour asymmetries can nonetheless
generate the observed baryon asymmetry,
due to time-dependent effects \cite{CJN}}.

The asymmetries given are for the 
decays of any of the    $\psi_I$.
To obtain
a large enough CP asymmetry, at least  two of the $\psi_I$
must be very degenerate :$M_I - M_J \sim
 \Gamma_I$, so the CP asymmetries in the
decays of at least two of the $\psi_I$s will
contribute  to the baryon asymmetry. It is
therefore  fortunate
that the CP asymmetries do not vanish
when summed on  $I$:  in
a two generation model,   in the limit
where $M_1 \to M_2$,  the CP asymmetries
are equal
$\epsilon^{(1, \ell)}  = \epsilon^{(2, \ell)}$,
rather than opposite.

As anticipated, the sum of
$\epsilon^{(I, \ell)}$ and $\epsilon^{(I, s)}$
vanishes because the model conserves lepton number.
It will therefore be important that  no interaction
be fast enough to equilibrate the singlet
and doublet sectors, as the asymmetries are
produced and until the Electroweak Phase Transition.
This is discussed in section \ref{sec:equil}.

It is helpful to be able to estimate the size
of these CP asymmetries. 
This scenario of out-of-equilibrium decay 
leptogenesis at the electroweak scale requires
a large CP asymmetry  $\epsilon$  $\gsim 0.1$. 
So from eqn (\ref{epsl}), 
all the elements 
$ [\lambda]_{I \a}$ and $[y]_{I \b}$  must be comparable,
for all $\a, I, \b$,
and the $\psi$s degenerate :$M_1 - M_2 \gsim \Gamma$.
For phases to maximise the asymmetry, large mixing angles, 
and comparable eigenvalues $\lambda_1 \sim \lambda_2$,
$y_1 \sim y_2$,
\beq
\epsilon^I
\sim 
\frac{1}{16 \pi}\frac{y_1^2 \lambda_1^2}{y_1^2 +2\lambda_1^2}
\frac{M_I}{M_I -M_J}  \leq \frac{1}{4D}  
\label{epsdev}
\eeq
where the last approximation expresses the mass splitting
in units of $\Gamma_I$
\beq 
M_I - M_J = D \Gamma_I
\label{defnD}
\eeq and the coupling constant combination was
maximised  by taking
\beq
y_1 \sim \sqrt{2} \,\lambda_1 ~~.
\label{sqrtdeux}
\eeq
 Recall that the formulae (\ref{epsl}) and 
 (\ref{epss})
are valid for  $M_I -M_J > \Gamma_{I}$, so 
$\epsilon \sim  0.1$ is barely consistent.
Since the mass $M$ is the only interaction linking
the singlet and doublet sector, 
loop corrections  should not destabilise the
small splitting.

\subsection{Washout of the asymmetry by inverse decays and scattering}
\label{sec:equil}

\begin{figure}[ht]
\begin{center}
\includegraphics[width=5cm]{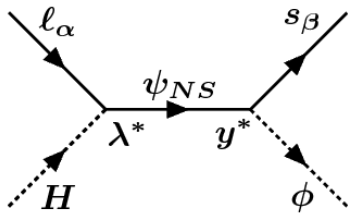}
\end{center}

\caption{
Dangerous scattering process $\ell + H \rightarrow \overline{s} + 
\overline{\phi}$
\label{fig:scat}} 
\end{figure}

As the $\psi$s decay, 
they produce an  asymmetry in doublet leptons. 
This asymmetry will survive,  and be  redistributed
throughout the plasma  by the interactions
in chemical equilbrium \cite{KSHT},
if  all the interactions which
can destroy   
this asymmetry   are out
of equilbrium. 
Such destructive interactions  are refered to as
 ``washout 
interactions'',
and in our model,  are the
inverse decays $ \ell H \to \psi$
and $\overline{\phi}\overline{s} \to \psi$,
which transfer lepton number back to
the $\psi$s, 
and the  $H \ell \to \overline{\phi}
\overline{s}$ scattering
(see   figure
\ref{fig:scat})  which 
exchanges lepton number between the doublets
and light singlets. 
We follow the  tradition\footnote{See
\cite{PRep} for a discussion, and references, of this curious
separation.} of treating 
the on-shell-$\psi$ part part of figure \ref{fig:scat} 
as inverse decays, and including the remaining
 `` Real Intermediate State -subtracted'' part as
scattering far
below the $\psi$ pole. 

For making estimates, it is convenient to have thermally
averaged rates, as opposed to the reaction densities
\footnote{see, for instance, \cite{PRep} for definitions.}
used in Boltzmann Equations. So we
define the thermally averaged inverse decay rates  as
\bea
\label{gaminv}
\langle \Gamma( \ell_\a H \to \psi_I) \rangle
&=&
\frac{ \gamma(\Psi_I \rightarrow \ell_\a H)}{n_\ell(T)} 
\simeq \frac{n_\psi^{(eq)} (T)}{n_\ell(T)}\Gamma(\Psi_I \rightarrow \ell_\a H) 
  \eea
where $n_\ell(T) \simeq g_\ell T^3/\pi^2$,
 the $\psi$ decay rates are given in eqn (\ref{Gamtree}),
 and   the equilibrium
abundance of a non-relativistic particle $X$ is
\beq 
\label{neq}
n_X^{eq}(T) \simeq g_X \left[ \frac{  m_X T}{2 \pi} \right]^{3/2} e^{-m_X/T}
\eeq
where $g_X$ is the number of spin degrees of freedom
of $X$ (notice that our number densities describe
particles, not the particles + anti-particles).
Comparing the inverse decay rate of
 eqn (\ref{gaminv}) and the annihilation
rate of   (\ref{gelann}) to the Hubble expansion, shows that
they freeze out at a similar temperature 
\beq
T_{BAU} \simeq \frac{M}{20}
\label{TBAU} 
\eeq
 The annihilations, discussed in section \ref{sec:DM}, 
will  freeze out first for
\beq
\label{Y4<}
 |[Y Y ^\dagger]_{II}|^2
 < \frac{{M}_\psi^3}{T^3} 2\pi^2 | y_{\b I}|^2,
4\pi^2 |\lambda_{\a I}|^2
~~~.
\eeq
We will see that this condition is satisfied,
so annihilations   stop depleting the
$\psi$ number density before    the
inverse decays allow a lepton asymmetry to survive.

The number density of $\psi$s remaining
when  both inverse decays
  $\langle \Gamma( \ell H \to \psi_I) \rangle $ and
$\langle \Gamma( \phi s \to \overline{\psi_I}) \rangle$
are out of equilibrium  is
\beq
n_\psi
(T_{BAU})  = {\rm min} \left\{
\frac{H(T_{BAU})}{ \Gamma (\Psi_I \rightarrow \ell H) } n_\ell(T_{BAU}) ~,~
\frac{H(T_{BAU})}{ \Gamma (\Psi_I \rightarrow \overline{\phi} 
\overline{s}) }n_s(T_{BAU}) \right\} ~~
\label{gelinv}
\eeq
where $\Gamma$ and $H$ are given in
equations (\ref{Gamtree}) and   (\ref{gelann}).

To protect a doublet lepton asymmetry, the
inverse decays from the singlet and doublet
sectors  must both  be out of equilibrium. To see this,
recall that the heavy Dirac  $\psi$s have $L = 1$, and there
are equal numbers of $\psi$s and $\overline{\psi}$s
to begin. As they decay,  opposite lepton
asymmetries develop in the doublet leptons and light
singlet sector. Consider now 
the case that inverse decays from the doublets 
 are out of equilibrium, but that
the asymmetry in light  singlets can be transfered
back to the heavy Dirac leptons by  $\phi s \to \overline{\psi_I}$.
Then the anti-asymmetry from the light singlet
sector is transfered  to the $\psi$s and
$\overline{\psi}$s, who transmit it to the
doublets via decays, which are always in equilibrium.

Away from the $\psi$ mass pole, the scattering
cross-section for $\ell + H \rightarrow \overline{s} + 
\overline{\phi}$ is
\beq 
\sigma(\ell_\a + H \rightarrow \overline{s}_\b + 
\overline{\phi}) =
\frac{|y_{\b I} \lambda_{\a I} |^2}{32 \pi} \frac{M_I^2}{(s-M_I^2)^2} 
\to \frac{|y_{\b I} \lambda_{\a I} |^2}{32 \pi M_I^2} 
\eeq
where,  after the arrow, 
 the internal line
momentum is neglected. This is an acceptable approximation
at temperatures $T \lsim M_I/20$,
  where the  inverse decays are out of equilibrium.
With this approximation, the thermally averaged
scattering rate $ \Gamma \sim \sigma n_\ell$
is out of equilibrium for
\beq 
|y_{\b I} \lambda_{\a I} |^2 \lsim 
4 \times 10^{-11} 
\frac{M_I}{3{\rm  TeV}} 
~~~.
\label{diffbd}
 \eeq

\subsection{Putting it all together in the Majorana case}

The ``washout'' interactions go out of equilibrium at
$T_{BAU}$ (given in eqn (\ref{TBAU})), after which  an
asymmetry in the SM leptons is  produced. This
should occur prior to the electroweak phase transition,
which we take to occur at $T_{EPT} \sim$ 150 GeV 
\cite{dORT}. Requiring  $T_{BAU} > T_{EPT} $,
imposes a lower bound on the singlet mass scale
\beq
M \gsim  3~ {\rm TeV}~~.
\eeq 
There is also an upper bound on $M$, as a function
of $\lambda$ and $Y$, from requiring
that the inverse seesaw give the observed neutrino
mass differences.  This condition was given in eqn
(\ref{mnunum}). Since a large enough CP asymmetry requires
large mixing angles and comparable eigenvalues in
$\lambda$ and  $y$, we can restrict to the
one generation case:
\beq
m_{atm}= 0.59
\left( \frac{ \lambda_2}{10^{-9}}\right)^2
\left( \frac{(3 ~{\rm TeV})}{ M_2}\right)^2
  \frac{ Y_2 \langle \phi \rangle}{v} \, {\rm  eV}
\label{matm}
\eeq
And finally, the requirement that  
$\psi \bar{\psi} \to \phi \bar{\phi}$ annihilations freeze-out 
before  the inverse decays $ \ell_\a H \to \psi$, 
imposes the relation between $Y$ and
$\lambda$ given in eqn  (\ref{Y4<}). Expressed in terms
of eigenvalues, we obtain 
\beq
Y_2^4 \lsim 10^{-3}\left(\frac{\lambda_2^2}{10^{-9}}\right) ~~.
\label{r3}
\eeq
    
\begin{figure}[ht]
\begin{center}
\includegraphics[width=10cm]{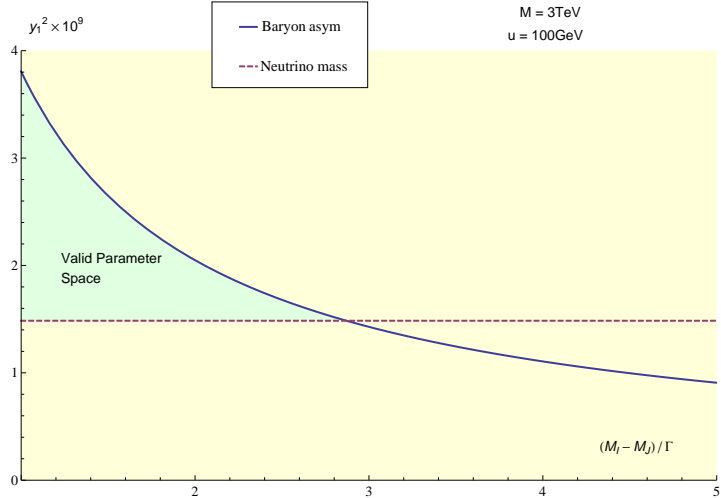}
\end{center}
\caption{Available parameter space  generating the observed
baryon asymmetry and neutrino masses, of Majorana type,
in the inverse seesaw model with extra light singlets,
described by the Lagrangian (\ref{L}). \label{fig:maj}}
\end{figure}

Electroweak sphalerons
will partially  transform   the  $B-L$ 
asymmetry produced in SM fermions,  $Y_{B-L}$,    to a baryon
asymmetry\cite{KSHT}
$Y_B = (12/37)Y_{B-L}$.
So the baryon asymmetry produced in our model will
be of order
\beq 
\label{dev}
Y_B \simeq \frac{12}{37}
  \frac{n_\ell - n_{\bar{\ell}}}{s}
\simeq  \frac{12}{37}  \frac{4 n_\psi(T_{BAU}) }{s} \, \epsilon
\simeq 4 \times 10^{-10}
 \frac{ M_\psi }{3~ {\rm TeV}} \frac{10^{-9}}{|\lambda_2|^2} 
  \frac{1}{D}~~,
\eeq
where  $4 n_\psi =  \Sigma_I (n_{\psi_I} +  n_{\bar{\psi}_I})$,
and we used eqns 
(\ref{gelinv}) and
(\ref{epsdev}).
It is marginally possible to obtain the
atmospheric mass difference and the observed baryon
asymmetry, with 
 $M = 3$ TeV,   $Y_2$  saturating eqn
(\ref{r3}),  and $\langle \phi \rangle \lsim v$. 
In figure \ref{fig:maj}, the available
parameter space is plotted as a function
 $|y_1|^2/ 10^{-9}$ in eqn (\ref{defnD}). 

\subsection{Dirac limit}
\label{deME}

As seen in the section above,
for  Majorana light neutrino masses,
 the parameter space of our model is very restricted. 
In this section, we explore the
Dirac limit  of the light neutrino mass matrix (\ref{mass}),
where all Majorana mass terms are zero, $\mu_A \to 0 ~~ , A \in \{ X, Y, Z\}$. 
Without Majorana mass terms one can redefine  the lepton number of 
$\{N^c, S,s,\phi\} $ to be $  \{ -1, 1, - 1, 0\}$ 
 so that also after the phase transition of $\phi$ 
lepton number is conserved. 
Neutrinos become Dirac particles with $s_I$ being their right 
handed Dirac partners. The light mass eigenvalue is 
\beq m_\nu ^D = \frac{m_D \mu_y}{M} \eeq 
which is only suppressed by $\frac{1}{M}$ compared to $\frac 1{M^2}$ in the Majorana case. 
This will significantly enhance our allowed parameter space. But it is also clear that hirarchy in neutrino masses must now come from a hierarchy in the Yukawa eigenvalues $\lambda_1 < \lambda_2$, and/or the same for $y$.

\begin{figure}[ht]
\begin{center}
\includegraphics[width=10cm]{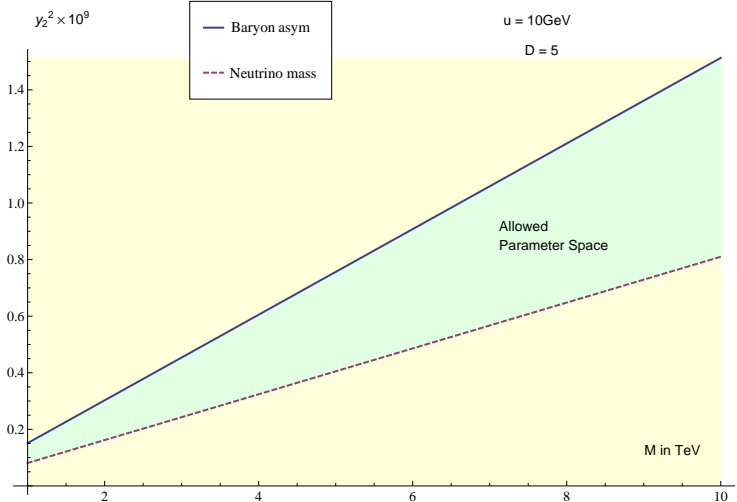}
\end{center}
\caption{Constraints from  the Baryon asymmetry 
and neutrino masses on $y_2^2$ as a 
function of the parent particle mass scale $M_\psi$ 
give the allowed parameter space in the Dirac
 limit. The upper bound from Baryon asymmetry 
(solid line) varies with the degeneracy 
 $M_I - M_J \approx D \Gamma(\psi_I \to all)$. 
For less degenerate parent particles, 
the slope reduces and the allowed parameter 
space shrinks. The dashed line is the 
lower bound from neutrino masses and 
varies with the vacuum expectation
value  $u$. For smaller $u$, the slope 
rises and the parameter space shrinks.}
	\label{fig:Dirac}
	\end{figure}

Before looking at the parameter space in the Dirac model, we shall look at its thermal 
history, comparing to the Majorana case. The first thing to realize is, that annihilation 
processes do not exist. The only processes that washout $CP$ asymmetry are scattering, 
as in fig.(\ref{fig:scat}) and inverse decays which, like in the Majorana case, 
determine the moment of efficient $CP$ violation. We
assume the inverse decay rates from the various flavours
are comparable, despite the mild hierarchy in $\lambda$ and/or $y$. 
After the  inverse decays  become 
slow, 
all heavy parent particles will decay and produce a $CP$ asymmetry that is not 
washed-out, so that the first part of equation (\ref{dev}) remains valid.
Therefore the thermal history in the Dirac limit does not differ 
significantly from the Majorana case.
 The value of $\epsilon$,  however, will
be reduced,  because of the Yukawa hierarchy. To be concrete,
in figure \ref{fig:Dirac}, we take  $\epsilon = \frac{1}{16 D}$ (compare
to the Majorana case, eqn  (\ref{epsdev})).

One should also  have a careful look 
at Big Bang Nucleosynthesis because it restricts
the number of thermalised neutrinos at $T \sim MeV$: $N_\nu \lsim 4$. It was 
shown in \cite{Shapiro} that right handed Dirac neutrinos  are not in equilibrium
 at $T\sim MeV$ because their interaction rates are suppressed by a factor $\left(\frac{m_\nu}{T}\right)^2$ 
compared to that of left handed neutrinos. Consequently the number of neutrinos 
in thermal equilibrium at BBN is not changed by our model with respect to the Standard Model.

Let us now turn to the parameter space of the Dirac limit. In the Majorana case the 
allowed parameter space is very small because asymmetry production favours small $y$, 
in order to produce a big density of $\psi$ when inverse decays freeze out which can 
then be transformed into enough baryon asymmetry, even if $\epsilon <1$. But neutrino 
masses prefer big $y$ couplings, because the neutrino mass term is already suppressed 
by $1/M^2$. This forces us to take quite large values for $u$, the vev of $\phi$ in order 
to enhance neutrino masses without limiting asymmetry production. In the Dirac limit neutrino 
masses are less suppressed, so that the conflict between baryon asymmetry and neutrino 
masses is less critical. We can realize our model with moderate values for $u \approx 6 GeV$, 
far away from electroweak phase transition, while parent masses are at  $M=3TeV$ and do not
 have to be more degenerate than $M_I - M_J \approx 5 \Gamma(\psi_I \to all)$.
 These values 
were obtained for a ideal constellation of our parameters, i.e. all phases in our Yukawa 
couplings $\lambda$ and $y$ are of order one and that   $2\lambda_2^2 \sim y_2^2$. 
However in this Dirac limit small variations from these ideal conditions are possible as the 
parameter space is big enough to compensate for non ideal conditions. A plot of possible 
parameter space is given in fig.(\ref{fig:Dirac}).

\section{Discussion}
\label{disc}

This paper  considered  baryogenesis  at the electroweak
scale, in an out-of-equilibrium decay scenario.   This was
implemented in  an inverse seesaw model  with
extra light  singlets $s$,  described by the
Lagrangian (\ref{L}).  The aim was to
relate the baryon asymmetry to the dark matter abundance,
and possibly fit various neutrino anomalies with
the extra singlets. Similar Dark Matter and baryon abundances
were supposed to arise, because, 
if the Dark Matter is made of WIMPs, and the baryon asymmetry
produced in electroweak-scale out-of-equilibrium decay,
then  both  relic densities are controlled
by  electroweak-scale interactions going  out of
equilibrium. The baryon asymmetry is also proportional
to a CP asymmetry $\epsilon$, see eqn (\ref{dev}).

The scenario for generating the baryon asymmetry in our
model is outlined at the end of the introduction. 
It does not work very well. The first
difficulty  is that it does not
naturally give  the large  CP asymmetry 
that is required,  
contrary to models with explicit lepton number
violation (reviewed  in section \ref{sec:models}).
More importantly,  the allowed parameter
space  where it  produces a sufficient baryon asymmetry
and correct neutrino masses is tiny
(see figure \ref{fig:maj}), and unattractive.
It requires
a small  mass splitting between  the heavy singlets
$M_2 - M_1 \simeq \Gamma$, but the formula
for $\epsilon$, eqn(\ref{epsl}), becomes unreliable
in this limit.   Also,  the  inverse
seesaw is attractive because  it allows  new LHC-scale
singlet fermions with large Yukawa couplings, and
spontaneous lepton number violation due to a small vev. 
However, our model requires  small yukawas $\lambda$
and $y$, to ensure that  interactions which wash out
the baryon asymmetry freeze out soon enough.  This
forces a large lepton number violating vev, of order the
Higgs vev. The masses of the extra light  singlets $s$ are arbitrary,
but would be $\lsim$ GeV, unless a steep
hierarchy is tuned into the couplings (see discussion after 
eqn (\ref{mnunum})). A larger
successful parameter space  is obtained (see figure \ref{fig:Dirac}), 
if the Majorana couplings of the  inverse seesaw model
are set to zero. Then 
 the active neutrinos obtain Dirac masses with
the light steriles. However, then the motivation
for the singlet
scalar (whose vev spontaneouly broke
lepton number in the Majorana case),  
is no longer clear. Since it no longer
carries lepton number, it could  mix
with the Standard Model Higgs, and  some analysis
would be required to determine  its
experimental signatures.

Despite the flaws in our model,  it could be
unsurprising, if not natural, to obtain
similar baryon and dark matter relic densities,
when the dark matter are WIMPs, and the baryon asymmetry
is produced in out-of-equilibrium decay. 
 In such scenarios, the baryon asymmetry
is  produced after the freeze-out of  washout interactions. 
This condition
can be roughly estimated as
\beq
\Gamma(all \to \psi) \simeq
\Gamma(\psi \to all) e^{-M/T} \lsim H(T_{BAU})
\eeq
where $\psi$ is the decaying parent of the BAU, of mass
$M$, and $H$ is the Hubble expansion rate. Notice
that inverse decays are Boltzmann-suppressed, because the light
decay products have difficulty to find, in
the thermal bath, the energy to produce a heavy $\psi$.

This Boltzmann suppression is reminiscent of the freezeout
of annihilations of a WIMP $\chi$, which can be estimated to occur
when 
\beq
n_{\chi} (T_{DM}) \langle \sigma v \rangle
\lsim H (T_{DM})
\eeq
The $e^{-m/T}$  appears here in
equilibrium number density of a non-relativistic
particle, see eqn  (\ref{neq}).

From these two estimates, it is straightforward to
estimate 
the ratio of co-moving densities
\footnote{Note that the 
WIMPs and BAU-parents may never  simultaneously
have this ratio of densities; 
the physically relevant ratio is $n_b/n_\chi$.}
of BAU parents $\psi$ to WIMPs $\chi$ as
\beq
\frac{Y_\psi}{Y_\chi}  \simeq \frac{1}{(2 \pi z)^{3/2}} \frac{g^4}{\lambda^2}
\frac{M_\psi}{m_\chi}
\sqrt{\frac{g(T_{DM})}{g(T_{BAU})}}
\label{*}
\eeq 
where the decay rates and annihiation cross-section
were normalised to the relevant particle masses
$\Gamma \equiv \lambda^2M_\psi/(8 \pi)$, and 
$\langle \sigma v \rangle \equiv g^4/(8 \pi m_\chi^2)$,
and we assume that the baryon-parents
and WIMPs freeze out at comparable values of
$z =m/T \simeq 25$.
At a given
mass scale, this shows that the co-moving number density of
BAU-parents can be  easily  of the same order
as that of  particles which annihilate.
So weak-scale BAU parents naturally have the
same abundance as weak-scale dark matter.

However, all BAU parents do not neccessarily produce baryons.
On average, a parent produces $\epsilon$ baryons,
where $\epsilon$ parametrises  CP violation, and
is suppressed by a loop factor, mixing angles and possibly small
couplings. This  could suggest 
that   $\epsilon$ is naturally $\lsim 10^{-3}$. 
Eqn (\ref{*}) predicts $$
\frac{n_B}{n_{DM}} \sim
\epsilon \frac{Y_\psi}{Y_\chi} 
 = C \epsilon  \frac{M_\psi}{ m_\chi}
$$
where the coefficient $C$  
could be ${\cal O}(1)$, for  $\lambda^2 \ll g^4$.
So to obtain  $n_B \sim n_{DM}$ 
requires $\epsilon \sim 1$ for $m_\chi \sim M_\psi$,
or $m_\chi < M_\psi$ for  $\epsilon < 1$.
It could be  reasonable to suppose that
 the WIMP is an electroweak-scale particle, with
gauge couplings $g$, and that  the baryon-parent is a flavour-scale
particle, with flavoured couplings $\lambda$, in which
case, such
a ratio of couplings could be credible.
Then for  ``natural'' values of
the CP asymmetry  $\epsilon \lsim 10^{-3}$, and  the  flavour-scale
$M_\psi \gsim 100$ TeV, the baryon and WIMP 
number densities today are  ``naturally'' comparable.

\section*{Acknowledgements}

We thank Steve Abel for participating in
the genesis of this project, Nuria Rius for discussions and 
important comments,  Christopher Smith
for many useful conversations, and Marco Cirelli for a relevant question.
S.D. and M.E.  acknowledge partial support from the  
European Union FP7  ITN INVISIBLES 
(Marie Curie Actions, PITN- GA-2011- 289442).

\end{document}